\documentstyle[prb,aps,twocolumn,psfig]{revtex}
\begin{document}

\draft

\title{The transverse magnetoresistance of the 
two-dimensional chiral metal}

\author{J. T. Chalker}

\address{Theoretical Physics, Oxford University,
1 Keble Road, Oxford OX1 3NP, 
United Kingdom}

\author{S. L. Sondhi}

\address{Department of Physics, Princeton University, Princeton NJ 08544, USA}
\date{\today}

\maketitle

\begin{abstract}
We consider the two-dimensional chiral metal, which exists at the surface
of a layered, three-dimensional 
sample exhibiting the integer quantum Hall effect.
We calculate its magnetoresistance in response to a component of magnetic 
field perpendicular to the sample surface, in the low temperature, but
macroscopic, regime where inelastic scattering may be neglected.
The magnetoresistance is positive, following a Drude form with a field scale,
$B_0=\Phi_0/al_{\text{el}}$, given by the transverse field strength at which 
one quantum of flux, $\Phi_0$, passes through a rectangle with sides set 
by the layer-spacing, $a$, and the elastic mean free path, $l_{\text{el}}$. 
Experimental measurement of this magnetoresistance may therefore
provide a direct determination of the elastic mean free path
in the chiral metal.

\end{abstract}
 

\section{INTRODUCTION}

The existence of a new electronic phase, 
the two-dimensional chiral metal, formed at the surface of 
a layered, three-dimensional quantum Hall conductor
has recently been predicted theoretically, 
\cite{chalkerdohmen,balentsfisher}
and confirmed experimentally. \cite{gwinn}
This chiral metal arises from hybridization of the edge states 
associated with each layer of the quantum Hall conductor.
The resulting surface phase is decoupled from states in the bulk of the 
conductor, which at the Fermi energy must be localized by disorder
for the Hall conductance to be quantized.
The distinguishing characteristics of the phase arise from the 
fact that electron motion is chiral along the edge of each layer,
but diffusive in the direction perpendicular to the layers.
In particular, drift in the chiral direction
prevents repeated, phase coherent scattering of an electron from
any given impurity, suppressing localization effects 
completely, outside the mesoscopic regime.
By contrast, interference effects in the mesoscopic regime 
have a variety of consequences, which have 
been investigated theoretically in some detail.
\cite{chalkerdohmen,balentsfisher,kim,mathur,yu,bfz,grs,cho,wang,wangplerou,meir}

Transport measurements are, of course, the obvious experimental
probe of the chiral metal.
A surface response current flowing in the chiral direction
is associated with the quantized Hall conductance of the 
three-dimensional system.
In constrast, surface 
conductivity in the non-chiral, 
diffusive direction depends on
microscopic parameters of the
system: the elastic mean free path, 
$l_{\text{el}}$, 
the inter-layer coupling energy, $t$, the chiral drift velocity, $v$, 
and the inelastic scattering rate, 
$\tau^{-1}_{\text{in}}$;
specifically, the mean conductance per
square, $\sigma$ is given \cite{balentsfisher} in units of $e^2/h$ by 
$\sigma = at^2l_{\text{el}}/(\hbar v)^2$. 
Knowledge of 
the amplitude of mesoscopic conductance fluctuations, together
with sample dimensions, should allow determination of the 
inelastic scattering length. \cite{cho}
The first detailed studies of this kind 
\cite{gwinn} involved vertical transport in 
multi-quantum well semiconductor samples; 
conduction by surface states also provides an
interpretation of 
earlier \cite{stormer}
and more recent \cite{brooks} experiments on the bulk quantum Hall effect
in semiconductor samples, as well as phenomena in organic metals 
in strong magnetic fields. \cite{organics}

In this paper we suggest that measurements of the 
transverse magnetoresistance of the chiral metal 
-- the dependence of surface resistance
on a magnetic field component, $B_{\perp}$, perpendicular
to the surface -- may be a useful source of additional information.
As anticipated by Balents and Fisher,\cite{balentsfisher}
this magnetoresistance is positive.
We calculate it in the (macrosccopic) low-temperature 
regime, in which the inelastic scattering rate is much smaller than that for 
elastic scattering, obtaining 
a Drude form
\begin{equation}
\label{result}
\sigma(B_{\perp})=\frac{\sigma(B_{\perp}=0)}{1+(B_{\perp}/B_0)^2}\,,
\end{equation}
with $B_0=\Phi_0/al_{\text{el}}$.
This result is exact for the model we study. Its simplicity is a 
direct consequence of the elimination of multiple scattering 
processes by chiral motion. Moreover, while the value of 
$\sigma(B_{\perp}=0)$ depends on two unknown microscopic quantities,
the elastic mean free path and the chiral velocity, 
the magnetoresistance field scale, $B_0$, involves only the first of these.
Studies of magnetoresistance are hence potentially 
both a test of our understanding of the chiral metal, and a way to determine
separately the values of $l_{\text{el}}$ and $v$.

We give a qualitative arguments that lead to our results
in section II, present a detailed calculation in 
section III, and add remarks bearing on experiment in section IV.

\section{MODEL AND QUALITATIVE DISCUSSION}

Consider first the edge of a single, two-dimensional layer
which has unit quantized Hall conductance.
Under the combined influence of the
magnetic field component normal to the layer
and the confining potential at the edge of the sample,
electrons at the Fermi energy
will drift along the edge, acquiring at most
a phase shift from impurity scattering or randomness in the 
position of the edge.
Within a single particle description,
this may be represented using the Hamiltonian
\begin{equation}
\label{1edge}
H=vp_x + V(x),
\end{equation}
where $p_x$ is the momentum operator in the 
direction parallel to the edge. The potential, $V(x)$,
includes impurity contributions, and also generates the Aharonov-Bohm
phase that electrons accumulate if their path wanders to enclose magnetic
flux because of surface roughness.

By extension,\cite{balentsfisher} 
the Schr\"odinger equation for a many-layer
sample with a surface in the $x-z$ plane,
hopping energy, $t$, between neighboring edges
(labeled by integer $n$ and separated with spacing $a$), 
and transverse magnetic field,
$B_{\perp}$, represented by the vector potential 
${\bf A}=(B_{\perp}an,0,0)$, is 
\begin{eqnarray}
\label{model}
\nonumber
(H\psi)_n(x)& = & v(-i\hbar \partial_x + eB_{\perp}an)\psi_n(x) \\ 
\nonumber
& &   -t [\psi_{n+1}(x) + \psi_{n-1}(x)]\\
& &  +V_n(x)\psi_n(x)\,.
 \end{eqnarray}
This is the model that we study in this paper.

There are two limitations of this model that are worth noting. Both
arise from restricting the single-edge problem to one dimension. In
doing so we have to worry about the fact that, in semiclassical terms,
the guiding center trajectories of particles with different energies 
will lie at different distances from the bulk of the sample,
and will therefore experience different impurity potentials. This has
the consequence that the scattering phase shift will depend upon the
energy of the state---such an effect is not representable in our
model as we will see below. A second, potentially more serious effect,
is the introduction of a random spatial ($x$) dependence into the 
$z$-axis hopping on account of the different wanderings of the uncoupled
edges in neigboring layers; this is representable, but not included in
our model. An extreme case would be the directed network model of 
Ref[1] which will lead to very different results as
we discuss in Section IV.

\begin{figure}
\centerline{\psfig{figure=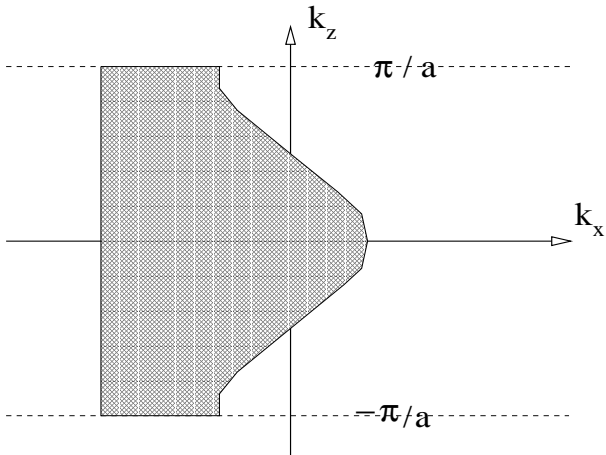,width=8cm,height=6cm}}
\caption{Fermi sea of chiral metal, with wavevectors $k_x$ in the
chiral direction and $k_z$ in the inter-layer direction.}
\label{fig1}
\end{figure}

Without impurity scattering or a transverse magnetic field,
the surface states fill a one-sided Fermi sea as sketched in Fig\,~1.
Impurity scattering generates a finite conductance 
\cite{chalkerdohmen,balentsfisher}
which can be estimated as follows.
First, note that eigenfunctions,
$\psi(x)$, of the single-edge Hamiltonian with energy $E$
have the form
\begin{equation}
\psi(x)=\psi(0)\exp(\frac{i}{\hbar v}
\int_0^x[ E-V(x') ]dx')\,.
\end{equation}
The phase acquired as a result of the disordered potential (which
is independent of energy as promised),
$\alpha(x)=(1/\hbar v)\int_0^x V(x')dx'$ is of no 
importance for a particle propagating along an isolated edge,
but sets the mean free path,
$l_{\text{el}}$, for a system of coupled edges,
this being the distance an electron must propagate
to accumulate a random phase
of unit magnitude. Suppose for definiteness that 
the potential on the $n^{th}$ edge, $V_n(x)$, 
is gaussian distributed with short-range correlations, so that 
$\langle V_n(x) \rangle = 0$ and 
$\langle V_n(x) V_m(x') \rangle = \Delta \delta_{nm}\delta(x-x')$.
Then the condition $\langle \alpha^2(x=l_{\text{el}}) \rangle =1$ results in
$l_{\text{el}}=(\hbar v)^2/\Delta$.

Given the mean free path, it is easy to
estimate the diffusion constant, $D$, and 
surface conductivity, $\sigma$, in the transverse direction.
Typical velocities in this direction 
have magnitude $at/\hbar$, while the scattering 
rate is $v/l_{\text{el}}$, so that
\begin{equation}
D \sim \left(\frac{at}{\hbar}\right)^2\frac{ l_{\text{el}}}{v}\,.
\end{equation}
The density of states in energy is $(h v)^{-1}$ per unit length
for a single edge, and $(h v a)^{-1}$ per unit 
area for the surface, and hence the Einstein relation 
gives
\begin{equation}
\sigma \sim \frac{e^2}{h} \frac{at^2l_{\text{el}}}{(\hbar v)^2}\,.
\end{equation}

Consider now the effect of a transverse magnetic field.
In the absence of impurities, the Lorentz force arising from
the chiral motion will sweep electrons across the Brillouin zone 
in the non-chiral direction, in a time $(h/a)/(eB_{\perp}v)$.
Correspondingly, electrons follow a snaking path in real space 
(Fig\,~2)
with an amplitude, $A$, for oscillations of their coordinate in the
interplane direction which decreases with increasing $B_{\perp}$,
having the dependence $A \sim t/eB_{\perp}v$.
The field scale for the magnetoresistance is the field strength
at which the period of these oscillations is comparable 
to the elastic scattering time, $l_{\text{el}}/v$,
from which follows the value $B_0 \sim h/eal_{\text{el}}$.
At field strengths much smaller than this, 
the electron scatters
too frequently from impurities 
for its path to be influenced by the transverse
magnetic field, while at much larger field
strengths its interplane coordinate
follows a random walk, with step length $A$ and step rate 
$v/l_{\text{el}}$. The resulting diffusion coefficient,
$D(B_{\perp}) \sim t^2/(e^2B^2_{\perp}vl_{\text{el}})$,
is reduced from its zero-field value by a factor
$(B_0/B_{\perp})^2$, as is therefore also the conductivity,
producing the large-field behavior of Eq(\ref{result}). 
\begin{figure}
\centerline{\psfig{figure=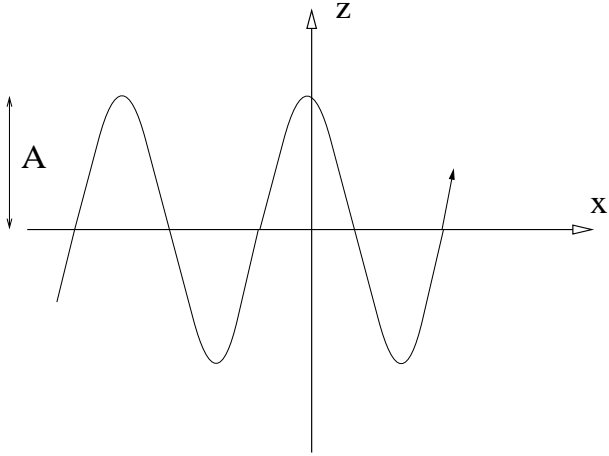,width=8cm,height=6cm}}
\caption{Trajectory of electrons in real space, in the presence of a transverse
component to the magnetic field, with coordinates $x$ in the chiral direction
and $z$ in the inter-layer direction, and oscillation amplitude $A$.}
\label{fig2}
\end{figure}

\section{Calculation}

The disorder-averaged one- and two-particle Green functions
can be calculated exactly for the chiral metal, provided disorder
correlations are short-range, because motion in the chiral
direction suppresses repeated scattering from any given impurity.
As a result, modeling the impurity potential with 
Gaussian white noise, the  one-particle 
Green's function is given {\it exactly}
by the Born approximation, and the two-particle function
by a sum of ladders. We show below that corrections 
to this behavior in a potential with a finite
correlation length, $\lambda$, are small in powers
of $\lambda/l_{\text{el}}$; they are also
strongly irrelevant (in the renormalization group sense) to 
long-distance properties, as shown by Balents and Fisher
\cite{balentsfisher}.

We simplify notation by
introducing the dimensionless field strength, $b=eB_{\perp}va/t$,
and choosing units in which $t=\hbar v =1$, so that
the model (Eq\,\ref{model})
takes the form
\begin{eqnarray}
\label{model2}
\nonumber
(H\psi)_n(x)& = & -i\partial_x\psi_n(x)
  - [\psi_{n+1}(x) + \psi_{n-1}(x)]\\
  \nonumber & & 
+bn \psi_n(x) +V_n(x)\psi_n(x)\\
 & \equiv & (H_0\psi)_n(x)+V_n(x)\psi_n(x)\,.
\end{eqnarray}
We use a position vector, ${\bf r}=(x,n)$ with one continuous and
one discrete component.
Denoting the Green's function by $g \equiv (z-H)^{-1}$,
we require its disorder average,
\begin{equation}
G(z;{\bf r}_1,{\bf r}_2) \equiv \langle g(z;{\bf r}_1,{\bf r}_2) \rangle\,,
\end{equation}
the diffusion propagator,
\begin{equation}
K(\omega;{\bf r}_1,{\bf r}_2) \equiv
\langle 
g(\omega+i0;{\bf r}_1,{\bf r}_2) 
g(-i0;{\bf r}_2,{\bf r}_1)\rangle\,,
\end{equation}
and in particular the behavior at small 
wavevectors of its Fourier transform,
\begin{equation}
K(\omega;{\bf k}) \equiv \int_{-\infty}^{\infty}dx\sum_n 
e^{i{\bf k}\cdot{\bf r}}K(\omega;{\bf 0},{\bf r})\,.
\end{equation}

A first step is to treat the problem without disorder.
The eigenfunctions of $H_0$ satisfy
\begin{equation}
(H_0\psi)_n(x) = (k + \epsilon_{\alpha} )\psi_n(x)
\end{equation}
with
\begin{equation}
\psi_n(x) = \frac{1}{\sqrt{2\pi}}e^{ikx}\phi_{\alpha}(n),
\end{equation}
where
\begin{equation}
\label{phi}
\epsilon_{\alpha}\phi_{\alpha}(n) = bn\phi_{\alpha}(n) -
\phi_{\alpha}(n+1) - \phi_{\alpha}(n-1)\,.
\end{equation}
This equation has the solution
\begin{equation}
\phi_{\alpha}(n)=J_{n-\alpha}(2/b)
\end{equation}
with $\epsilon_{\alpha}=\alpha \cdot b$ and $\alpha$ integer,
where $J_{l}(x)$ is the Bessel function of order $l$.

The single-particle Green function 
in the absence of disorder, $g_0(z)\equiv (z-H_0)^{-1}$,
is given by
\begin{equation}
\nonumber
g_0(z;{\bf r}_1,{\bf r}_2)= 
\frac{1}{2\pi}\int_{-\infty}^{\infty}dk \sum_{\alpha}
\frac{e^{ik(x_2-x_1)}\phi_{\alpha}(n)\phi_{\alpha}(m)}
{z-(k+\epsilon_{\alpha})}
\end{equation}
In particular, for ${\cal I}m(z) >0$, 
$g_0(z;{\bf r}_1,{\bf r}_2)=0$ if $x_2 < x_1$, while for
$x_2 = x_1 +0^+$, $g_0(z;{\bf r}_1,{\bf r}_2)=-i\delta_{nm}$.

Disorder generates a self-energy after averaging, which
is diagonal in real space and given in the Born
approximation by
\begin{equation}
\Sigma(z) = -i\frac{\Delta}{2}{\rm sgn}[{\cal I}m(z)]\,.
\end{equation}
In order to examine corrections to the Born approximation,
suppose temporarily that the 
Gaussian distributed potential, $V_n(x)$, is piecewise
constant on segments of length $\lambda$, so that
the second cumulant is
$\langle V_n(x) V_m(x') \rangle = (\Delta/\lambda) \delta_{n,m}$,
if there is an integer $N$ for which
$N\lambda<x,x'<(N+1)\lambda$, and zero otherwise.
Self energy contributions at $p^{th}$ order are proportional to
$(\Delta/\lambda)^p \lambda^{2p}(g_0)^{2p-1}$, where the factor of 
$\lambda^{2p}$ arises from integration over 
internal position coordinates, restricted to an interval
of range $\lambda$. Restoring dimensionful units,
the $
p^{th}$ order contribution is proportional to 
$(\lambda/l_{\text{el}})^p\cdot (g_0)^{-1}$, so that only 
the Born term ($p=1$) need be retained as 
$(\lambda/l_{\text{el}}) \to 0$.
Hence we have
\begin{equation}
\label{g}
G(E+i0;{\bf r}_1,{\bf r}_2) 
=\frac{1}{2\pi}\int_{-\infty}^{\infty}\!\!\!\!dk \sum_{\alpha}
\frac{e^{ik(x_2-x_1)}\phi_{\alpha}(n)\phi_{\alpha}(m)}
{E+i\Delta/2-(k+\epsilon_{\alpha})}\,.
\end{equation}

Analogous arguments show that the 
diffusion propagator is given in the same limit by a sum 
of ladder diagrams, with the result that
\begin{equation}
K(\omega;{\bf k}) =\frac{\Lambda(\omega;{\bf k})}
{1-\Delta \Lambda(\omega;{\bf k})}\,,
\end{equation}
where
\begin{equation}
\Lambda(\omega;{\bf k}) =
\int_{-\infty}^{\infty}dx\sum_n 
e^{i{\bf k}\cdot{\bf r}}G(\omega+i0;{\bf 0},{\bf r})G(-i0;{\bf r},{\bf 0}).
\end{equation}
Using Eq.(\ref{phi}) and (\ref{g}), we obtain
\begin{equation}
\Lambda(\omega;{\bf k}) = \sum_{l}\frac{J_{l}^2(4\sin(k_y/2)/b)}
{\Delta-i(k_x+\omega+ lb)}\,.
\end{equation}
At small wavevectors this has the behavior
\begin{equation}
K(\omega;{\bf k})=\frac{1}{i(\omega +k_x) +Dk_y^2}
\end{equation}
with
\begin{equation}
D=\frac{2\Delta}{\Delta^2+b^2}.
\end{equation}
Using the Einstein relation to obtain the conductivity,
 we arrive at our main result: Eq (\ref{result})\cite{fn-hall}.

\section{Discussion}

The observability of transverse magnetoresistance 
depends on a number of factors. The inelastic scattering rate, 
$\tau_{\rm in}^{-1}$, must be smaller than the interchain hopping rate,
$t/\hbar$, for transport not simply to be incoherent; and for
the theory we have described to apply,
$\tau_{\rm in}^{-1}$  should also be smaller than 
the elastic scattering rate, $v/l_{\rm el}$.
The fact that mesoscopic conductance fluctuations have been observed
\cite{gwinn} in
transport by surface states in semiconductor samples
seems a good indication that 
both these conditions 
can be met experimentally at low temperature.
If this is the case, the remaining condition
is that the field scale, $B_{0}$,
should not be too large. 
In order that $B_{0}$ is much smaller than 
the magnetic field component parallel to the sample surface,
which is responsible for the bulk quantum Hall effect, we require 
a clean, flat surface, so that 
the elastic mean free path satisfies
$l_{\text{el}} \gg l_{\rm B}^2/a$, where $l_{\rm B}$ is the magnetic length 
within the layers of the sample, and $a$ is the inter-layer spacing.

We should remark that the intrinsic magnetoresistance we have
discussed is distinct from the simple reduction of the
inter-plane matrix element by
an in-plane magnetic field, analyzed for example in Ref[\onlinecite{macd}].
The dependence of the tunneling energy, $t$, on $B_{\perp}$
will make a contribution to the magnetoresistance that is
additional to the one we calculate, and with a field scale that will be
much larger than $B_0$ if $l_{\rm el} \gg l_{\rm B}$.
The two contributions should be distinguishable by comparing
surface magnetoresistances in two configurations.
In one, the magnetic field component that lies within the plane of
the sample layers is directed
normal to the sample surface; in the other,
it lies within the surface plane.
A magnetic field normal to the surface should generate a magnetoresistance
via both mechanisms, whereas a magnetic field within the surface plane
will affect only the inter-layer tunneling energy. A rectangular sample
with a large aspect ratio would be ideal for such a comparison.

Finally, as noted in the discussion following Eq (3), substantial roughness
at the surface may imply that something like the directed network model 
of Ref [1] is a better description of the details of transport at the edge.
This would lead to a second type of chiral metal, which would exhibit {\it no}
intrinsic magnetoresistance, even in the absence of inelastic scattering.
In such a system, inter-layer tunneling takes place only at discrete
points, and not continuously in the chiral direction as in
Eq~(\ref{model}). If a large random scattering phase
is accumulated between tunneling points, then a transverse magnetic field 
component can have no effect. Consequently, a null result for the 
magnetoresistance would plausibly indicate that strong spatial randomness
in the $z-$axis tunneling is essential for a proper description of the
transport in experimental systems.

\section*{acknowledgements}
We are grateful to the Institute for Theoretical Physics, UCSB, for hospitality
during the completion of this work, which was supported in part by
NSF Grants No. PHY94-07194 (ITP), DMR-9632690 (SLS), the A. P. Sloan 
Foundation (SLS), and by EPSRC grant GR/J8327 (JTC).

\end{document}